# Nonlocal Friction Forces in the Particle-Plate and Plate-Plate Configurations: Nonretarded Approximation

G. V. Dedkov and A. A. Kyasov

Nanoscale Physics Group, Kabardino-Balkarian State University, Nalchik, Russia

In the nonrelativistic approximation of fluctuation electrodynamics, using the specular reflection model and the nonlocal dielectric permittivity of a metal, we obtained simple analytical expressions for the friction forces in the particle-plate and plate-plate systems upon relative motion of the bodies with constant velocity. It is shown that at separations of about $1 \div 10\ nm$, for an Au nanoparticle (or a gold plate) moving near another gold plate at rest, the dissipative forces are 2 to 4 orders of magnitude higher than in the case when the local Drude dielectric permittivity is used.

## 1. Introduction

A rigorous description of the van der Waals friction (quantum friction) and dissipative effects in nanostructures is of great fundamental and practical importance in connection with the intensive development of micro- and nanotechnology, since these effects may affect the behavior of micromechanical devices (MEMS).[1] However, until recently, theoretical calculations of these effects caused a lot of discussion even for the simplest particle – plate and plate – plate configurations (see Refs. 2–4 and the corresponding references). In contrast to the measurements of the static conservative van der Waals and Casimir-Lifshitz forces,[5,6] experimental measurements of the relevant dissipative (frictional) forces are still at the initial stage. In the latter case, as shown in Refs. 6,7, the role of the spatial dispersion (SD) effect is relatively small, but in the case of quantum friction, spatial dispersion has a much stronger influence on the magnitude of the friction force.[8,9] This is, in particular, due to the possibility of generating electron-hole pairs when the interacting bodies are in relative motion even at a low velocity[8].

The aim of this work is to calculate the van der Waals friction force and quantum friction force with a more detailed justification of the approach based on the use of the specular reflection model (SRM)[10,11] and the nonrelativistic approximation ($c \to \infty$) of fluctuation electrodynamics in the case of arbitrary temperatures and direction of motion of a small particle relative to a thick plate. In particular, it is known that the SRM model reproduces many properties of real surfaces and has been successfully used to calculate the energy loss of energetic ions near a solid surface and the dynamics of electron motion in the wake potential.[12]

Based on these results, the case of interaction of two plates in relative motion is treated using the "correspondence principle" between the particle –plate and plate –plate configurations.[13] Unlike,[8,9] where the authors explored the quantum friction between two plates and an atom with a surface at $T = 0$, the use of the realistic analytical approximation for the nonlocal dielectric function of the metal enables performing analytical and numerical calculations of the quantum friction force and van der Waals friction force at an arbitrary temperature $T$ in a much simpler way.

## 2. Theoretical Formulation

Scheme of the SRM and the motion of a particle with a fluctuation dipole moment **d**($t$) relative to the surface of a polarizable plate is shown in Fig. 1. In the vacuum region $z > 0$, the electric field is created by the dipole **d**($t$) with the Cartesian components $(V_x, 0, -V_z)$ of the velocity vector **V**, and the identical dipole with components $(V_x, 0, -V_z)$ of the velocity vector **V'**, mirror-symmetric relative to the plane $z = 0$, and a fictitious charge density $\rho_s(x, y, t)\delta(z)$ on the plane $z = 0$.

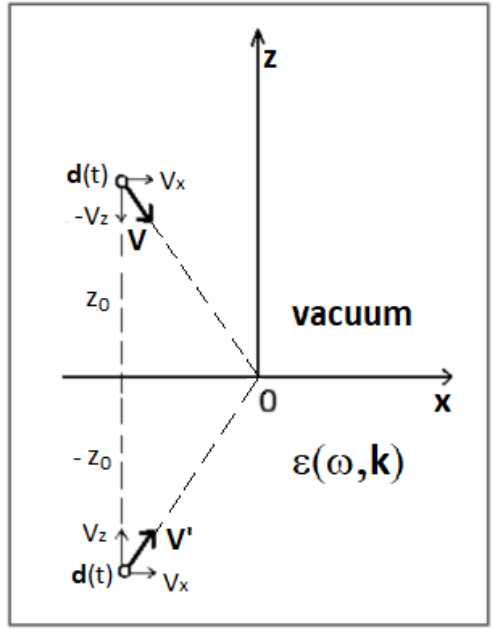

Fig. 1. Scheme of the particle motion and coordinate system used.

The field inside the plate ($z < 0$) is also created by the charge density $\rho_s(x,y,t)\delta(z)$. By setting the polarization vectors of dipole sources in the form $\mathbf{P} = \mathbf{d}(t)\delta(\mathbf{r} - \mathbf{V}t)$ and $\mathbf{P}' = \mathbf{d}(t)\delta(\mathbf{r} - \mathbf{V}'t)$, decomposing electric potential and corresponding charge densities $-\text{div}\mathbf{P}, -\text{div}\mathbf{P}', \rho_s(x,y,t)\delta(z)$ into Fourier integrals with respect to frequency $\omega$ and three-dimensional wave vector $\mathbf{k} = (\mathbf{q}, k_z) = (k_x, k_y, k_z)$, solution of the Poisson equations $\Delta\Phi = -4\pi\rho$ for the Fourier components of the potential at $z > 0$ and $z < 0$ can be written as

$$\Phi(\omega,\mathbf{k}) = -\frac{4\pi}{k^2}\left[i\mathbf{k}\mathbf{d}(\omega - \mathbf{k}\mathbf{V}) + i\mathbf{k}\mathbf{d}(\omega - \mathbf{k}\mathbf{V}') + \rho_s(\omega,q)\right], \; z > 0. \tag{1}$$

$$\Phi(\omega,\mathbf{k}) = \frac{4\pi\rho_s(\omega,q)}{k^2\varepsilon(\omega,\mathbf{k})}, \quad z < 0. \tag{2}$$

where $\mathbf{d}(\omega - \mathbf{k}\mathbf{V})$, $\mathbf{d}(\omega - \mathbf{k}\mathbf{V}')$ and $\rho_s(\omega,q)$ are the Fourier-components of the dipole moments and $\rho_s(x,y,t)$. The condition of continuity of the potential $\Phi$ at the boundary $z = 0$ determines the quantity $\rho_s(\omega,q)$:

$$\rho_s(\omega,q) = -\frac{q}{\pi + qI_0}\int_{-\infty}^{\infty}\frac{dk_z}{q^2 + k_z^2}\left[i\mathbf{k}\mathbf{d}(\omega - q_xV_x + k_zV_z) + i\mathbf{k}\mathbf{d}(\omega - q_xV_x - k_zV_z)\right] =$$

$$= -\frac{\pi}{\pi + qI_0}\left[i\mathbf{q}\mathbf{d}(\omega^+ - q_xV_x) + i\mathbf{q}\mathbf{d}(\omega^- - q_xV_x) + qd_z(\omega^+ - q_xV_x) + qd_z(\omega^- - q_xV_x)\right] \tag{3}$$

$$I_0 = \int_0^\infty \frac{dk_z}{k^2\varepsilon(\omega,\mathbf{k})}, \tag{4}$$

where $\omega^\pm = \omega \pm iqV_z$. When calculating integral (3), it is considered as the limit at $z \to 0$ from the same expression with additional factor $\exp(i\omega k_z)$ in the integrand, and the integration over $k_z$ is performed along a contour including the real axis and the upper semicircle of the complex plane. In addition, we use the modification $\mathbf{k}\mathbf{d} = \mathbf{q}\mathbf{d} - k_zd_z$ of the scalar product,[14] which is necessary for matching the subsequent results to the exact solution for the potential obtained using the local dielectric permittivity. The boundary condition for the continuity of induction at the boundary $z = 0$ is satisfied automatically, since the component of potential (1) with the Fourier component due to the contribution of the dipoles is symmetric with respect to the boundary $z = 0$ and has a zero derivative with respect to the coordinate $z$, and the contributions to the induction from fictitious charge density are dropped out at $z \to \pm 0$. In the framework of the SRM, any charge distribution associated with a moving particle has the same property. The Fourier component $\Phi^{ind}(\omega,\mathbf{k})$ of the potential induced in the vacuum region is found from $\Phi(\omega,\mathbf{k})$ by subtracting the Fourier component $\Phi^{vac}(\omega,\mathbf{k})$ of the potential for a particle moving in vacuum. In order to do this, we should make the replacement $\varepsilon(\omega, \mathbf{k}) \to 1$ in (1)–(4). As a result, we obtain

$$\Phi^{ind}(\omega,\mathbf{k}) = \frac{2\pi}{k^2}\Delta(\omega, q) \cdot [i\mathbf{q}\mathbf{d}(\omega^+ - q_xV_x) + i\mathbf{q}\mathbf{d}(\omega^- - q_xV_x) + qd_z(\omega^+ - q_xV_x) + qd_z(\omega^- - q_xV_x)], \quad (5)$$

$$\Delta(\omega, q) = \frac{\pi - qI_0}{\pi + qI_0}. \quad (6)$$

Taking into account (5), the expression for the induced potential at the point $(\mathbf{r}, t) = (\mathbf{R}, z, t) = (x, y, z, t)$ in the vacuum region takes the form

$$\Phi^{ind}(\mathbf{R}, z, t) = \frac{1}{(2\pi)^4}\int d\omega d^2q dk_z \Phi^{ind}(\omega,\mathbf{k})\exp(i(\mathbf{qR} + k_z z - \omega t)). \quad (7)$$

When a particle moves parallel to the surface, substituting (5) into (7), one needs to use the limit $V_z \to 0, V_x \to V$ under the condition $V_z t \to \pm z_0$, where $z_0$ is the particle distance from the surface. After that, integrating (7) by $k_z$ in the same way as in (3), we obtain the Fourier component of the potential with decomposition by frequency and two-dimensional wave vector

$$\Phi^{ind}(\omega, q, z) = \frac{2\pi}{q}\Delta(\omega, q)e^{-q(z+z_0)}\left[iq_x d_x(\omega - q_xV) + iq_y d_y(\omega - q_xV) + qd_z(\omega - q_xV)\right]. \quad (8)$$

Formula (8) coincides with the exact solution of the electrodynamic problem in the case of parallel motion, when the local dielectric permittivity $\varepsilon(\omega)$ is used.[14] Using (8), further calculation of the dissipative van der Waals force acting on the particle completely repeats the calculation with local $\varepsilon(\omega)$.[3,14] Expressing the dielectric response of the particle via the frequency-dependent polarizability $\alpha(\omega)$, we obtain the following result for $F_x$ ( $T_1$ and $T_2$ are the particle and plate temperatures in the units of energy, $\omega_+ = \omega + q_xV$)

$$F_x = -\frac{\hbar}{\pi^2}\int_0^\infty \int_{-\infty}^{+\infty} dq_x q_x \int_{-\infty}^{+\infty} dq_y q e^{-2qz_0}\text{Im}\Delta(\omega, q)\text{Im}\alpha(\omega_+)[\coth(\hbar\omega/2T_2) - \coth(\hbar\omega_+/2T_1)] \quad (9)$$

In the linear approximation in velocity, for $T_1 = T_2 = T$, from (9) it follows

$$F_x = -\frac{1}{2\pi}\left(\frac{\hbar^2 V}{T}\right)\int_0^\infty d\omega \int_0^\infty dq q^4 e^{-2qz_0}\text{Im}\Delta(\omega, q)\text{Im}\alpha(\omega)\sinh^{-2}(\hbar\omega/2T) \quad (10)$$

and in the limit of quantum friction ($T = 0$), respectively,

$$F_x = \frac{4\hbar}{\pi^2}\int_0^\infty dq_x q_x \int_0^\infty dq_y q e^{-2qz_0}\int_0^{q_xV} d\omega \text{Im}\alpha(\omega - q_xV)\text{Im}\Delta(\omega, q). \quad (11)$$

Formulas (9)–(11), quite naturally, completely coincide with the known results in the case of the local dielectric permittivity of the plate material. [2-4]

3. Spherical particle above metal surface

For the practical calculation of the force $F_x$ by formulas (10), (11), we use the expression for the polarizability $\alpha(\omega) = R^3(\varepsilon(\omega) - 1)/(\varepsilon(\omega) + 2)$ of a spherical particle with radius $R$ and the Drude dielectric constant $\varepsilon(\omega) = 1 - \omega_p^2/\omega(\omega + i\gamma)$, where $\omega_p$ and $\gamma$ are the plasma frequency and the damping factor ($\omega_p = 9\ eV$ and $\gamma = 30\ meV$ for gold). For the function $\varepsilon(\omega,\mathbf{k})$ of the metal plate, we use the well-known approximation,[15] which takes into account the linear-frequency asymptotic behavior of the Lindhard permittivity, the generation of electron-hole pairs and plasmons:

$$\varepsilon(\omega,\mathbf{k}) = 1 - \frac{\omega_p^2}{\{s^2 V_F^2[1-i\pi\omega\theta(2k_F-k)/2kV_F]-\omega(\omega+i\gamma)\}} \qquad (12)$$

where $k_F$ is the Fermi wave vector, $s = V_F/3$, $V_F$ is the Fermi velocity. In this case, the low-frequency expansion of the integral (4) leads to the expression $I_0 = A + B + C$, where

$$A = \pi k_{TF}^{-1}(1+x^2)^{-3/2}, \qquad (13)$$

$$B = -i\left(\frac{\pi}{k_{TF}}\right)\left(\frac{\omega}{\omega_p}\right)\left[\frac{(2+x^2)}{2(1+x^2)^{\frac{3}{2}}}\ln\left(\frac{p\sqrt{1+x^2}+\sqrt{p^2-x^2}}{p\sqrt{1+x^2}-\sqrt{p^2-x^2}}\right) - \frac{p\sqrt{p^2-x^2}}{(1+p^2)\sqrt{1+x^2}}\right]\theta(p-x), \qquad (14)$$

$$C = -i\left(\frac{\pi}{k_{TF}}\right)\left(\frac{\gamma\omega}{\omega_p^2}\right)\left[\frac{\sqrt{1+x^2}-x}{x\sqrt{1+x^2}} - \frac{1}{2(1+x^2)^{3/2}}\right]. \qquad (15)$$

Here $x = q/k_{TF}$, $k_{TF} = \sqrt{3}\omega_p/V_F$ is the inverse Thomas-Fermi screening length, $\theta(x)$ is the unit Heaviside function, $p = 2k_F/k_{TF} = \sqrt[3]{3\pi^2/2}\sqrt{a_B/r_s}$, $a_B$ and $r_s$ are the Bohr radius and the jellium parameter ($r_s/a_B = 3.01$ and $p = 1.415$ for gold). We also mean one and the same value $\gamma$ in $\varepsilon(\omega)$ and in (12). It should be noted that Eq. (14) is somewhat different from that presented in Ref. 11 by the expression in square brackets and the presence of factor $\theta(p-x)$. With allowance for (6) and (13)–(15) we obtain

$$\Delta(\omega,q) = \frac{1-(\omega/\omega_p)^2 S^2(x) + 2i(\omega/\omega_p)S(x)\sqrt{1+x^2}}{\left(\sqrt{1+x^2}+x\right)^2+(\omega/\omega_p)^2 S^2(x)}, \qquad (16)$$

$$S(x) = \left[\frac{x(2+x^2)}{2(1+x^2)}\ln\left(\frac{p\sqrt{1+x^2}+\sqrt{p^2-x^2}}{p\sqrt{1+x^2}-\sqrt{p^2-x^2}}\right) - \frac{xp\sqrt{p^2-x^2}}{(1+p^2)\sqrt{1+x^2}}\right]\theta(p-x) +$$

$$+\beta\left[\sqrt{1+x^2}-x-\frac{2}{2(1+x^2)}\right]. \qquad (17)$$

In (17), and hereinafter, we omit the small argument $\beta = \gamma/\omega_p$ in writing $S(x)$ for simplicity. In the case of low particle velocities $V \ll V_F$ and $S(x) \sim 1$ ($V_F$ is the Fermi velocity), the terms proportional to $\omega^2$ (14) can be ignored, since only low frequencies $\omega \ll \omega_p$ contributes to integrals (10), (11). Similarly, for $\text{Im}\alpha$ one can use a simpler expression $\text{Im}\alpha \approx 3i\gamma\omega/\omega_p^2$. Then, as a result of integration over the frequency $\omega$, formulas (10), (11) are reduced to

$$F_x^{(nl)} = -\frac{\pi}{8}\left(\frac{R}{z_0}\right)^3 \frac{\hbar\gamma}{z_0^2}\left(\frac{T}{\hbar\omega_p}\right)^2 \frac{V}{\omega_p} f_1(2k_{TF}z_0) \qquad (18)$$

$$F_x^{(nl)} = -\frac{1}{256\pi}\left(\frac{R}{z_0}\right)^3 \frac{\hbar\gamma}{z_0}\left(\frac{V}{\omega_p z_0}\right)^3 f_2(2k_{TF}z_0) \qquad (19)$$

where functions $f_i(x), i = 1,2$ have the form

$$f_1(x) = \int_0^\infty dz\, z^4 e^{-z} S_1(z/x), \qquad (20)$$

$$f_2(x) = \int_0^\infty dz\, z^6 e^{-z} S_1(z/x), \qquad (21)$$

$$S_1(x) = S(x) \frac{\sqrt{1+x^2}}{(\sqrt{1+x^2}+x)^2} \ . \tag{22}$$

If the local Drude function is used in calculating $F_x$, then $\Delta(\omega, q) \to (\varepsilon(\omega) - 1)/(\varepsilon(\omega) + 1)$ and performing elementary integration in (10), (11) we obtain

$$F_x^{(l)} = -3\pi \left(\frac{R}{z_0}\right)^3 \frac{\hbar\gamma}{z_0^2} \left(\frac{T}{\hbar\omega_p}\right)^2 \frac{V}{\omega_p} \frac{\gamma}{\omega_p} \tag{23}$$

$$F_x^{(l)} = -\frac{45}{16\pi} \left(\frac{R}{z_0}\right)^3 \left(\frac{T}{\omega_p z_0}\right)^3 \frac{\hbar\gamma}{z_0} \frac{\gamma}{\omega_p} \tag{24}$$

Comparing the fractions $F_x^{(nl)}/F_x^{(l)}$ corresponding to (18), (23) and (19), (24), we see that the increase in the friction force due to SD is mainly due to the presence of the large parameter $1/\beta = \omega_p/\gamma \gg 1$. This is in agreement with Ref. 9.

Figure 2 shows the calculated ratio $F_x^{(nl)}/F_x^{(l)}$ depending on $2k_{TF}z_0$ for the force of quantum friction (dashed line) and friction at a finite temperature (solid line). It is worth noting that in the case of gold we have $2k_{TF}z_0 = 34$ at $z_0 = 1\ nm$. As follows from Fig. 2, the effect of SD leads to an increase in the dissipative van der Waals force by more than two orders of magnitude at $z_0 = 1 \div 10\ nm$.

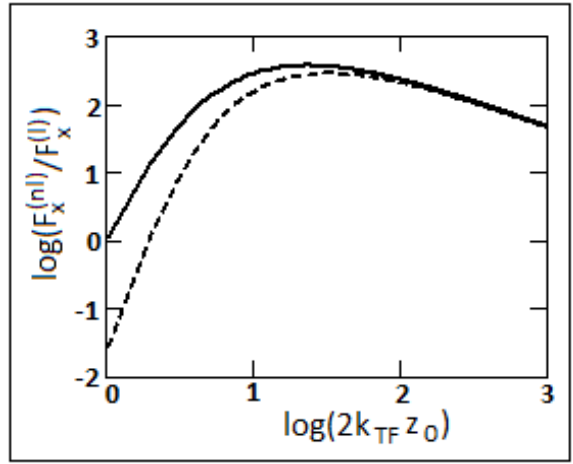

Fig. 2. Dependence $F_x^{(nl)}/F_x^{(l)}$ as a function of reduced distance $2k_{TF}z_0$ for an Au nanoparticle above the surface of gold. Dashed line: quantum friction force, solid line: friction force at a finite temperature; (nl) –nonlocal approximation (Eqs. (18), (23)), (l) –local approximation (Eqs. (19), (24)). The log functions hereinafter are with basis 10. For gold, $2k_{TF}z_0 = 34$ at $z_0 = 1\ nm$.

## 4. Two plates in relative motion

As Lifshitz first showed,[16] there is a simple "rule-of-thumb" between plate-plate (1) and particle-plate (2) configurations. This enables calculating the Casimir-Polder force in configuration 2 using the corresponding expression for the force in configuration 1. The rule reads

$$\Delta(\omega) \to 2\pi n \alpha(\omega) \tag{25}$$

$$F^{(2)}(z) = -\frac{1}{nS} \frac{d}{dl} F^{(1)}(l)_{l=z} \tag{26}$$

where $\alpha(\omega)$ is the polarizability of a particle of a rarified medium with density $n$ of particles that models the material of one of the plates, and $S$ is the surface area of the plates in the vacuum contact.

In systems outside of thermal and dynamic equilibrium, in the nonretarded limit, the corresponding transition rule was established in Refs. 3, 13. It was proved that relations (25), (26) are valid for all other quantities describing the fluctuation-electromagnetic interaction, such as the rate of heat exchange and friction force. Such a correspondence rule, obviously, must be valid in the nonlocal case too, for a limiting transition from the nonlocal case to the local one to exist. In addition, relations (25), (26) can be applied for both transitions: from configuration 1 to configuration 2 and vice versa.

With this in mind, to take into account SD in configuration 1, one should use the replacement $\Delta(\omega) \rightarrow \Delta(\omega, q)$ in (9). It is worth noting that the above approach is fundamentally different from that used in the calculations of the Casimir-Lifshitz attraction and friction forces with allowance for SD,[4,6] where the authors used the nonlocal Lindhard- Mermin dielectric permittivity $\varepsilon(\omega, \mathbf{k})$ (Ref. 17) in the expressions for the amplitudes of reflection of electromagnetic waves.

In our case, the friction force acting on the moving plate 1 per unit surface of the vacuum contact is formally the same as in Ref. 3 provided that $\Delta_1(\omega_+) \rightarrow \Delta_1(\omega_+, q)$ and $\omega_+ = \omega + q_x V$:

$$F_x(l) = -\frac{\hbar}{4\pi^3} \int_0^\infty d\omega \int d^2 q q_x e^{-2ql} \text{Im}\Delta_1(\omega_+, q) \text{Im}\Delta_2(\omega, q) |D|^{-2} \cdot$$

$$\cdot [\coth(\hbar\omega/2T_2) - \coth(\hbar\omega_+/2T_1)] \qquad (27)$$

where $D = 1 - \Delta_1(\omega_+, q)\Delta_2(\omega, q)\exp(-2ql)$ and $l$ is the spacing between interacting plates. Using (27), the expressions for the friction force in a linear approximation in velocity (at $T_1 = T_2 = T$) and in the case of quantum friction ($T = 0$) take the form ($n(\omega) = 1/(\exp(\hbar\omega/T) - 1)$ is the Planck factor)

$$F_x = \frac{\hbar V}{2\pi^3} \int_0^\infty d\omega \frac{dn(\omega)}{d\omega} \int_0^\infty d^2 q q_x^2 e^{-2ql} \text{Im}\Delta_1(\omega, q) \text{Im}\Delta_2(\omega, q) |D|^{-2}, \qquad (28)$$

$$F_x = \frac{\hbar}{\pi^3} \int_0^{+\infty} dq_y \int_0^\infty dq_x q_x e^{-2qz_0} \int_0^{q_x V} d\omega \text{Im}\Delta_1(\omega - q_x V, q) \text{Im}\Delta_2(\omega, q) |D|^{-2}. \qquad (29)$$

It is worth noting that in contrast to (27) and (29), the D-factor in (28) reads $D = 1 - \Delta_1(\omega, q)\Delta_2(\omega, q)\exp(-2ql)$. Formulas (28), (29) agree with those obtained by many authors[2-4,14,18] using local $\varepsilon(\omega)$, while (29) fully coincides with Eq. (24) in Ref. 8, obtained in the framework of the nonlocal quantum field theory.

As in Sec. 3, in the case under consideration ($V \ll V_F$), we can omit the terms proportional to $\omega^2$ in Eq. (16). Then $\Delta_{1,2}(\omega, q) \cong 1$ and $\text{Im}\Delta_{1,2}(\omega, q) \cong 2(\omega/\omega_p)S_1(x)$, and substituting these relations into (28), (28) yields (assuming that both plates are of the same material)

i)      $T_1 = T_2 = T$

$$F_x^{(nl)} = -\frac{1}{24}\left(\frac{T}{\hbar\omega_p}\right)^2 \left(\frac{\hbar V}{l^4}\right) f_3(2k_{TF}l). \qquad (30)$$

ii)      $T_1 = T_2 = 0$

$$F_x^{(nl)} = -\frac{1}{2^{11}\cdot 3}\left(\frac{V}{\omega_p l}\right)^3 \left(\frac{\hbar\omega_p}{l^3}\right) f_4(2k_{TF}l). \qquad (31)$$

$$f_3(x) = \int_0^\infty dy\, y^3 e^{-y} \frac{S_1^2(y/x)\left(\sqrt{1+(y/x)^2}+(y/x)\right)^4}{\left[\left(\sqrt{1+(y/x)^2}+(y/x)\right)^4 - e^{-y}\right]^2}. \tag{32}$$

$$f_4(x) = \int_0^\infty dy\, y^5 e^{-y} \frac{S_1^2(y/x)\left(\sqrt{1+(y/x)^2}+(y/x)\right)^4}{\left[\left(\sqrt{1+(y/x)^2}+(y/x)\right)^4 - e^{-y}\right]^2}. \tag{33}$$

On the other hand using the local Drude function $\varepsilon(\omega) = 1 - \omega_p^2/\omega(\omega + i\gamma)$, we have $\Delta_{1,2}(\omega) \cong 1$, $\mathrm{Im}\Delta_{1,2}(\omega) \cong 2(\omega\gamma/\omega_p^2)$. Substituting these relations into (28), (29) yields

$$F_x^{(l)} = -\frac{\zeta(3)}{4}\left(\frac{T}{\hbar\omega_p}\right)^2 \left(\frac{\hbar V}{l^4}\right)\left(\frac{\gamma}{\omega_p}\right)^2, \tag{34}$$

$$F_x^{(l)} = -\frac{5\zeta(5)}{2^8}\left(\frac{V}{\omega_p l}\right)^3 \left(\frac{\hbar\omega_p}{l^3}\right)\left(\frac{\gamma}{\omega_p}\right)^2. \tag{35}$$

where $\zeta(3) = 1.202$ and $\zeta(5) = 1.037$ are the Riemann zeta-functions.

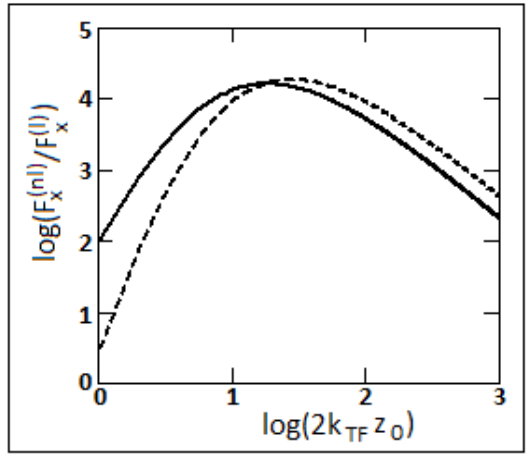

Fig. 3. Dependence $F_x^{(nl)}/F_x^{(l)}$ as a function of reduced distance $2k_{TF}l$ for two plates of gold in relative motion Dashed line: quantum friction force, solid line: friction force at a finite temperature; (nl) –nonlocal approximation (Eqs. (30), (34)), (l) –local approximation (Eqs. (31), (35)).

Shown in Fig. 3 is the calculated ratio $F_x^{(nl)}/F_x^{(l)}$ depending on $2k_{TF}l$, for two gold plates in relative motion. One can see that Fr in the range $l = 1 \div 10$ nm, friction forces increase by four orders of magnitude if SD is accounted for. The effect is due to the large factor $\omega_p^2/\gamma^2$.

## 5. Discussion

It is expedient to perform a more thorough comparison of the numerical values of the forces $F_x$ obtained using different approaches. Figure 4 shows a linear in velocity friction force between two plates of gold, as a function of separation $l$ for $T_1 = T_2 = 300\,K, V = 1\,m/s$.

Bold solid curve in Fig. 4 shows the nonlocal approximation, Eq. (30). The dashed thick curve and the curve shown by dots (S-local Drude and P-local Drude) show the results with local Drude function, performed using a recently transformed formula of the Levin-Polevoy-Rytov theory[19] (Eq. (A2) in the Appendix, Ref. 20). These curves correspond to the contributions of the second and first terms in square brackets of (A2). It is worth noting that in (A2), both evanescent and propagating modes are accounted for, though at short separations $l$ under consideration, the propagating modes make a small contribution to the result. The curve shown by long dashes in Fig. 4 was calculated by Eq. (34) and corresponds to the nonretarded local Drude approximation. We see that SD has a great effect only at short separations $l < 2\, nm$, whereas at larger $l$ the local Drude approximation due to the modes with S- polarization is dominant. All the curves in Fig. 4 obeys the law $F_x \sim V$, but the $l$−dependence is different: $F_x \sim l^{-5}$ for P-nonlocal Drude, $F_x \sim l^{-0.9}$ for S-local Drude, whereas for other two curves we have $F_x \sim l^{-4}$.

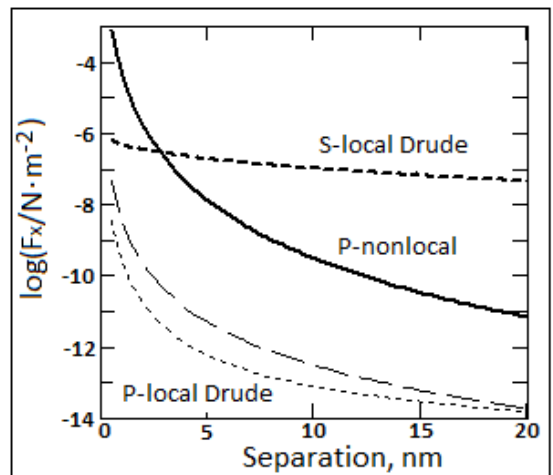

Fig. 4 Friction force between two gold plates at $T_1 = T_2 = 300\, K, V = 1\, m/c$. Curve P-nonlocal: Eq. (30); curve S-local Drude: Eq. (A2), the second term (the contribution from S-waves); curve P-local Drude: Eq. (A2), the first term (the contribution from P-waves); thin dashed line: Eq. (35).

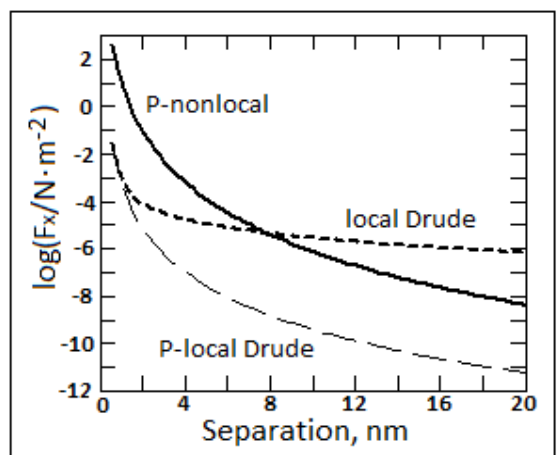

Fig. 5 Quantum friction force for two flat gold plates in parallel motion as a function of separation $l$ at $V = 0.1 V_F$ ($V_F = 1.4 \cdot 10^6\, m/s$). Thick solid curve (P-nonlocal): Eq. (31), curve local Drude: Eq. (A3) (summary contribution from S-waves and P-waves), thin dashed curve (P-local Drude) corresponds to the contribution from P-waves in (A3). The curve calculated by Eq. (35) (nonretarded local Drude for P-waves) is very close to P-local Drude curve and is not shown.

Figure 5 compares the quantum friction forces at $T = 0$. The bold solid curve was calculated by Eq. (31). The thick dashed curve (local Drude) was calculated by Eq. (A3) with account of both P-waves and S-waves. Curve P-local Drude corresponds to the contribution from P-waves only (Eq. (A3), the first term). The values of the force calculated by Eq. (35) nearly coincide with those for P-local Drude curve, and the results are not shown. We see that the range of distances with strong SD effect ($l < 8\ nm$) is greater than in the case of "thermal friction" (Fig. 4). The solid curve (P-local Drude) obeys the law $F_x \sim V^3/l^{7.2}$ at $2 < l < 20\ nm$ and $F_x \sim V^3/l^2$ at $1 < l < 2nm$. The local Drude curve obeys the law $F_x \sim V^3/l^{3.2}$ at $1 < l < 2\ nm$ and $F_x \sim V^3/l^2$ at $2 < l < 20\ nm$.

We also compared the results with Ref. 8, using two reference points in Fig. 8(a,b) of this paper corresponding to $V/V_F = 0.2$ and separation distances of 1.27 nm (24 $a_0$) and 5.3 nm (100 $a_0$). The jellium parameter ($r_s = 3\ a_0$) in Ref. 8 is close to our's ($r_s = 3.01\ a_0$). According to these data, $F_x \approx 9\ N/m^2$ and $F_x \approx 0.004\ N/m^2$, whereas Eq. (31) yields $F_x = 10.8\ N/m^2$ and $F_x \approx 0.0055\ N/m^2$, i.e. the matching is very good.

Despite the "positive" influence of SD at small separations, the absolute values of the friction forces are still very small. For example, in an ideal sphere-plane configuration, $F_{sp} \approx \pi R a F_z(a,V)$, where $R$ and $a$ are the curvature radius of the probing tip and the minimum separation distance, even at $R = 100 \mu m$, $V = 300\ m/s$ ($T = 300K$) we find by Eq. (30) $F_{sp} = 3 \cdot 10^{-15} N$ at $F_z = 0.01\ N/m^2$. In turn, in typical experimental situation, when the AFM technique is used, even this rather "moderate" velocity value requires a very large product of oscillation amplitude $A$ and frequency $f$ ($2\pi A f = 300\ m/s$). The high values of $A, R$, along with the small separation distance, make such an experiment very difficult. Probing the quantum friction force is still more difficult, since at $V = 300\ m/s$ we obtain a much lower value $F_z = 6 \cdot 10^{-8}\ N/m^2$, or we should increase the velocity to about $10^5 m/s$.

Finally, it is worthwhile to discuss the role of temperature. According to Eq. (30), the friction force scales as $F_x \sim T^2$ and decreases with decreasing temperature. Since the dissipation mechanism at SD is mostly due to the electron-hole excitation, this is quite natural. As for the general local theory,[20] our recent calculations showed a sharp growth in the friction force at $T < 50\ K$, when the damping factor $\gamma$ in the Drude formula for $\varepsilon(\omega)$ was changing according to the Blokh-Gruneisen law, $\gamma \sim T^5$. Therefore, the low-temperature behavior of the friction force is an intriguing issue and should be studied more thoroughly.

## 6. Conclusion

In this paper, we provide a derivation of the van der Waals friction force on a particle (thick metal plate) moving at constant velocity parallel to another thick metal plate (configurations 1 and 2). The theory is a generalization of the specular reflection model and the fluctuation-electromagnetic theory with account of the spatial dispersion of the material of the plate. Using the analytical approximation for bulk dielectric permittivity $\boldsymbol{\varepsilon}(\boldsymbol{\omega}, \mathbf{k})$ of metal, we obtained closed analytical expressions for the friction forces at a finite temperature $\boldsymbol{T}$ and in the case of quantum friction ($\boldsymbol{T = 0}$). When obtaining expressions for the friction force in the configuration of parallel plates, we used the principle of correspondence between configurations 1 and 2. This provides an unambiguous limit transition between formulas obtained in local and nonlocal theory.

Comparison of nonlocal and local formulations shows that the van der Waals friction force calculated using the nonlocal approach is 2–4 orders of magnitude higher than using the local approach (both for quantum friction and friction at a finite temperature). Mathematically, the increase in friction is due to the large factor $\omega_P/\gamma$. This is in agreement with Refs. 8,9. Moreover, a comparison of the numerical values of the nonlocal quantum friction force with those obtained in Ref. 8 within the quantum-field theory formalism, shows a very good coincidence.

The main theoretical result of this paper is that it provides a simple analytical approach for calculating the van der Waals friction force at separations between the bodies of order several nm, using a model expression for the dielectric permittivity of materials, which takes into account the frequency and spatial dispersion.

**Appendix A**

In Ref. 20, based on the original Rytov-Levin-Polevoy theory with local $\varepsilon(\omega)$,[19] we obtained a general expression for the friction force in configuration of two parallel thick plates with account of retardation (the corresponding quantities relating to plates 1 and 2 have the subscripts 1,2) at $V/c \ll 1$

$$F_x = -\frac{\hbar}{4\pi^3}\int_0^\infty d\omega \int d^2k\, k_x \left[\mathrm{Im}\left(\frac{q_1}{\varepsilon_1}\right)\mathrm{Im}\left(\frac{\widetilde{q_2}}{\widetilde{\varepsilon_2}}\right)\frac{|q|^2}{|Q_\varepsilon|^2} + \mathrm{Im}\left(\frac{q_1}{\mu_1}\right)\mathrm{Im}\left(\frac{\widetilde{q_2}}{\widetilde{\mu_2}}\right)\frac{|q|^2}{|Q_\mu|^2}\right] \cdot \quad (A1)$$

$$\cdot [\coth(\hbar\omega_-/2T_2) - \coth(\hbar\omega/2T_1)]$$

where $q = \sqrt{k^2 - (\omega/c)^2}$, $q_1 = \sqrt{k^2 - \varepsilon_1\mu_1(\omega/c)^2}$, $q_2 = \sqrt{k^2 - \varepsilon_2\mu_2(\omega/c)^2}$, the tilde means that the corresponding arguments of $\varepsilon_{1,2}$ and $\mu_{1,2}$ (dielectric permittivities and magnetic permeabilities) are taken at $\omega_- = \omega - k_x V$. Eq. (A1) describes the friction force acting on plate 1 that moves in the $x$-direction with constant velocity $V$. Moreover, in (A1) we have $Q_\varepsilon = (q + q_1/\varepsilon_1)(q + \widetilde{q_2}/\widetilde{\varepsilon_2})\exp(ql) - (q - q_1/\varepsilon_1)(q - \widetilde{q_2}/\widetilde{\varepsilon_2})\exp(ql)$, and $Q_\mu$ is described by the same expression with the change $\varepsilon \to \mu$. An important feature of (A1) is that it does not use the reflection factors, and the contributions from evanescent and propagating modes are not separated. Using (A1), the first-order-velocity approximation at $T_1 = T_2 = T$ yields (Eq.23) in Ref. 20)

$$F_x = \frac{\hbar V}{2\pi^2}\int_0^\infty d\omega\, \frac{dn(\omega)}{d\omega}\int d^2k\, k_x\left[\mathrm{Im}\left(\frac{q_1}{\varepsilon_1}\right)\mathrm{Im}\left(\frac{q_2}{\varepsilon_2}\right)\frac{|q|^2}{|Q_\varepsilon|^2} + \mathrm{Im}\left(\frac{q_1}{\mu_1}\right)\mathrm{Im}\left(\frac{q_2}{\mu_2}\right)\frac{|q|^2}{|Q_\mu|^2}\right] \quad (A2)$$

In the case $T_1 = T_2 = 0$, from (A1) we obtain the quantum friction force[20]

$$F_x = \frac{\hbar}{4\pi^3}\int_{-\infty}^{+\infty}dk_y\int_0^\infty dk_x\, k_x\int_0^{k_x V}\exp(-2kl)\mathrm{Im}\Delta_{1\varepsilon}\mathrm{Im}\widetilde{\Delta_{2\varepsilon}}\,|D_\varepsilon|^{-2} + (\varepsilon \to \mu). \quad (A3)$$

Here $\Delta_{i\varepsilon} = (\varepsilon_i q_i - q_i)/(\varepsilon_i q_i + q_i)$, $i = 1,2$, $D_\varepsilon = 1 - \Delta_{1\varepsilon}\widetilde{\Delta_{2\varepsilon}}\exp(-2kl)$, and the same expressions are used for magnetic contribution $(\varepsilon \to \mu)$. Note that in this paper we consider the case of nonmagnetic materials with $\mu_{1,2} = 1$.